\documentclass[10pt]{article}

\usepackage{lmodern}
\usepackage[utf8]{inputenc}
\usepackage[T1]{fontenc}
\usepackage{authblk} 
\usepackage{geometry}
\usepackage{titlesec}
\titleformat{\section}{\large\bfseries}{\thesection}{1em}{}
\titleformat{\subsection}{\normalsize\bfseries}{\thesubsection}{1em}{}
\usepackage{graphicx}
\usepackage{multicol}
\usepackage{url}
\usepackage{ragged2e}
\usepackage{amsmath}
\usepackage[authoryear]{natbib}
\bibpunct{(}{)}{;}{a}{}{,}

\usepackage{enumitem}
\usepackage{tabularx}
\usepackage{float}

\usepackage[most]{tcolorbox}

\newtcolorbox{greybox}{
  colback=gray!10,
  colframe=gray!50,
  boxrule=0.4pt,
  arc=1mm,
  left=7pt,
  right=7pt,
  top=7pt,
  bottom=7pt
}

\title{Methods for pitch analysis in contemporary popular music: Vitalic's use of tones that do not operate on the principle of acoustic resonance}

\author[1,3]{Emmanuel Deruty}
\author[2]{Pascal Arbez-Nicolas}
\author[3]{David Meredith}

\affil[1]{Sony Computer Science Laboratories, Paris, France, \texttt{emmanuel.deruty@sony.com}}
\affil[2]{Citizen Records, Dijon, France}
\affil[3]{Department of Architecture, Design and Media Technology, Aalborg University, Denmark}

\date{}

\begin{document}

\maketitle

\begin{greybox}
\small
Corresponding Medium page with embedded videos: \\\url{https://medium.com/@derutycsl/}\\\url{inharmonic-tones-and-listener-perception-in-vitalics-no-fun-978965a355d3}
\end{greybox}


\begin{center}

\justifying

\vspace{1em}
\begin{abstract}
Vitalic is an electronic music producer who has been active since 2001. Vitalic's 2005 track `No Fun' features a main synthesiser part built from a sequence of single inharmonic tones that evoke two simultaneous melodies. This part serves as a starting point for examining Vitalic's use of tones that do not operate on the principle of acoustic resonance. The study considers tones that evoke two or more simultaneous pitches and examines various inharmonic partial layouts. Examples outside Vitalic's music are also provided to suggest that similar tone properties can be found elsewhere in contemporary popular music.
\end{abstract}

\end{center}
\noindent





\section{Introduction}\label{sec:introduction}

Most musical instruments used in Western classical music operate on the principle of acoustic resonance. These instruments produce harmonic or quasi-harmonic complex tones, with components at frequencies that are integer multiples of a fundamental frequency. The perceived pitch approximates this frequency ($f_0$). Models for harmonic tones also specify a decrease in partial energy with increasing frequency \citep[p.~137]{mauch2010approximate}. Given a note on a musical score, a performer produces a tone with a fundamental frequency close to the frequency of the notated pitch. Conversely, in a music recording, pitch tracking processes are formulated as \emph{$f_0$ estimation tasks} \citep{drugman2018traditional,kim2018crepe}.



\begin{figure}
\includegraphics[width=\textwidth]{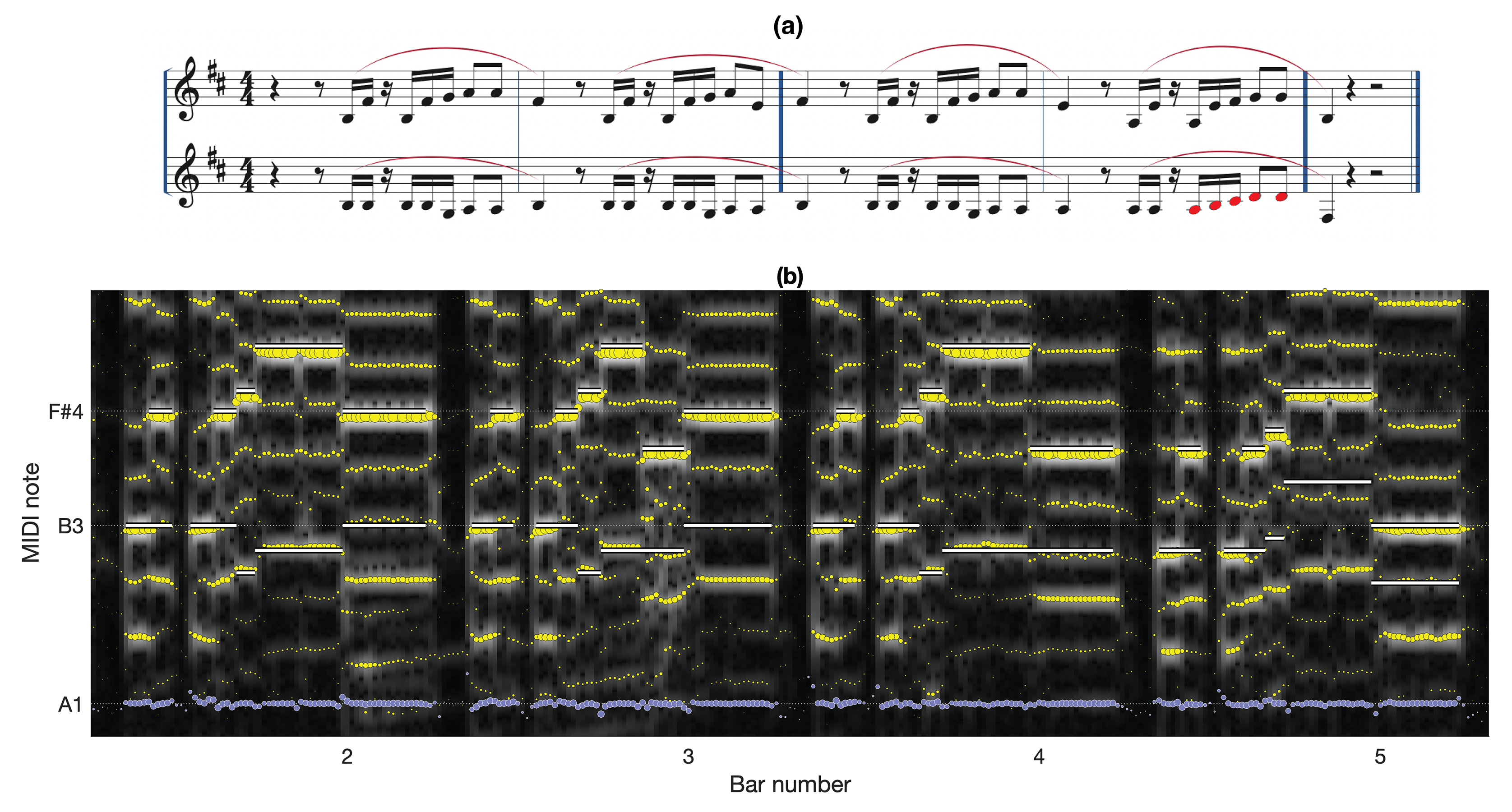}
\caption{`No Fun', 0'25 to 0'33. See suppl. mat. video
1 for the same figure with synchronized audio. (a) Approximate transcription. The red notes are difficult to hear. (b) Corresponding audio, STFT. The yellow scatter plot follows the partials. The blue scatter plot shows the median of the frequency difference between consecutive partials (27 first partials), weighted by the partials' energy. There is no partial at the frequency difference between consecutive partials. The white lines show the transcribed notes.} \label{fig:NoFunWideScore}
\end{figure}

In the music of French electronic producer Vitalic, different associations between tones and perceived pitches can be observed. An example is Vitalic's `No Fun' \citep{vitalic2005nofun}, for which Figure~\ref{fig:NoFunWideScore} (a) presents an approximate four-bar transcription from the song's main synthesizer track. Based on these perceived pitches, a reference to instruments operating under the principle of acoustic resonance would suggest two quasi-harmonic tones with simultaneous fundamental frequencies reflecting the transcribed melodic lines. However, the analysis shown in Figure~\ref{fig:NoFunWideScore} (b) reveals a sequence of single tones. Melodic movements correspond to changes in partial amplitudes and positions, with frequency differences between consecutive partials remaining close to a single value throughout the extract. The tones' partial amplitudes do not decrease as frequency increases. Section~\ref{sub:noisyshifted} provides more details on this type of tones, and Section~\ref{sec:freedom} more details on this synthesizer part.

The synthesizer part in Vitalic's `No Fun' exemplifies how tones not bound by the acoustic resonance principle allow for varied expressions of pitch. This paper aims to illustrate key aspects of such tones in Vitalic's music. It is part of a series, `Methods for pitch analysis in contemporary popular music'.

The remainder of the paper is structured as follows. Section~\ref{sec:background} introduces the methodological and theoretical frameworks. Section~\ref{sec:singletone} illustrates how a single tone, whether inharmonic or not, may produce one or several pitches that do(es) not correspond to the $f_0$. Section~\ref{sec:inharmonicity} proposes a typology of inharmonic tones in Vitalic's music. Section~\ref{sec:freedom} elaborates on `No Fun''s synthesiser part. Finally, Section~\ref{sec:others} provides examples that do not originate from Vitalic's music, suggesting that the use of tones not based on the acoustic resonance principle is not unique to Vitalic's music. Supplementary material with audio examples and detailed analyses is available at:

 \url{https://vitalic3-suppl-mat.s3.eu-west-3.amazonaws.com/B_index.html}

\section{Background}\label{sec:background}

\subsection{Vitalic}\label{sec:Vitalic}

Vitalic is the stage name of French electronic music producer Pascal Arbez-Nicolas, who has been active since 2001. For a comprehensive biography, see \citet{phares2024vitalic}. Vitalic cites Daft Punk and Giorgio Moroder as influences. According to informal interviews with the artist, Vitalic's style combines two complementary approaches. On the one hand, he enjoys making `dancefloor' tunes---dry, rhythmic pieces without many notes, such as `And it Goes Like' \citep{vitalic2023anditgoeslike}. On the other hand, he also likes to work on more `nostalgic' songs with `more notes' (his words), such as `Eternity' \citep{vitalic2017eternity}. Vitalic's discography is a mix of the two influences.

\newpage

\subsection{Terminology}

Regarding tones, we use the following vocabulary:

\begin{enumerate}

   \item \emph{Partial}: the spectral representation of a single periodic sine wave, corresponding to what \citet[p.~23]{helmoltz1885sensations} describes as a \emph{constituent} or \emph{partial tone}.

\item \emph{Harmonic complex tone}: an ensemble of partials that are integer multiples of a fundamental frequency ($f_0$), as defined by \citet[p.~23]{helmoltz1885sensations} as a \emph{compound}. When the partial corresponding to $f_0$ is absent, it can still be determined as the greatest common divisor of the partials' frequencies.

\item \textit{Complex tone} (or \textit{tone}): defined to be an ensemble of discrete partials that share a `common fate' \citep{Wertheimer1938}. Typically, the frequencies and/or amplitudes of these partials change in parallel---for instance, when the partials start and stop simultaneously. Common fate as a characteristic of tones was noted by \citet[p.~7]{rasch1982perception}. This definition permits some ambiguity, although no examples in this paper raise questions about tone identification. Such \textit{tones} may evoke more than one pitch, making the convention incompatible with \citet{moore1986thresholds}'s, for which a tone corresponds to one single pitch. Additionally, while the definition may encompass tones that do not evoke pitch, all examples in this paper are associated with pitch perception.

\item \emph{Inharmonic tone}: a tone that is not harmonic. An inharmonic tone can still display regularity in terms of where its partials are positioned. In this paper, all tones are inharmonic but exhibit such regularity---according to Vitalic, this is a result of preset selection during production. Following \citet[p.~9]{rasch1982perception}, `$f_0$' is used in the case of inharmonic tones, with $f_0$ evaluated as the fundamental frequency of the least deviating harmonic series.

\item \emph{Overtone} or \emph{upper partial}: refers to any partial in a tone other than $f_0$. This aligns with \citet[p.~23]{helmoltz1885sensations}'s definition of \emph{upper partial tones}.

\item \textit{Harmonics}: defined by \citet[p.~23]{helmoltz1885sensations} as the sinusoidal components of a harmonic complex tone, where the $n^{th}$ harmonic has a frequency $n$ times the fundamental. We extend the concept to inharmonic tones with regular partial positioning, where the $n^{th}$ harmonic's frequency is near the $n^{th}$ multiple of the $f_0$ of the least deviating harmonic series.

\end{enumerate}


\subsection{Pitch perception: spectral and temporal modeling}\label{subsec:modeling}

\citet{yost2009pitch} distinguishes between \textit{spectral} and \textit{temporal} modeling of pitch perception, whose main proponents are \citet{goldstein1973optimum} and \citet{licklider1951duplex}, respectively. In spectral modeling, pitch is extracted from the spectral components \textit{resolved} by the auditory periphery, i.e., whose frequency difference is above the critical bands' width. In temporal modeling, pitch is typically derived from autocorrelation. In this paper, \textit{spectral modeling} is used in a simplified form to link perceived pitches with individually audible upper partials. While \textit{temporal modeling} typically involves the highest autocorrelation peak excluding zero lag \citep[p.~401]{rabiner1976comparative}, we prefer analyzing frequency differences between consecutive partials. 

\begin{figure}[H]
  \centering
  \includegraphics[width=\textwidth]{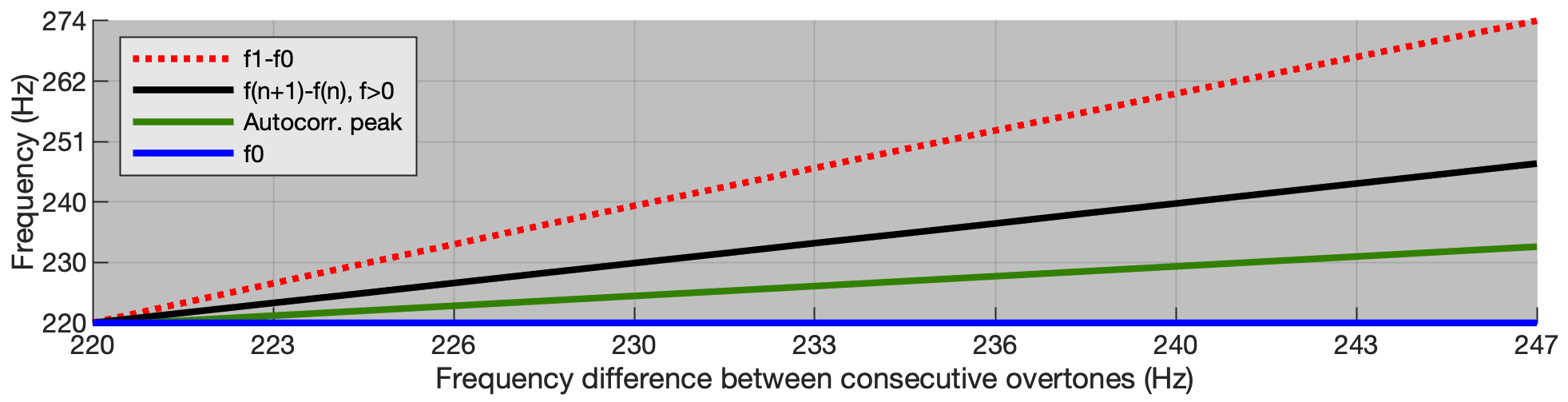}
  \caption{$F_0$, highest autocorrelation peak (excluding
zero lag), frequency difference between overtones, and frequency difference between the first two partials. The x-scale shows the frequency of which the overtones are multiple. The leftmost tone is harmonic, $f_0 = 220\text{Hz}$.\\}
\label{fig:inharmevolution}
\end{figure}

The reason for this choice is as follows. For harmonic tones, the highest autocorrelation peak (excluding zero lag) generally matches the frequency difference between partials. However, when tones are inharmonic, the highest autocorrelation peak (excluding zero lag) may not correspond to any perceived value. For example, Figure~\ref{fig:inharmevolution} compares the highest autocorrelation peak (excluding zero lag) with the frequency difference between partials for a series of increasingly inharmonic tones. In these tones, the $f_0$ remains at 220Hz (A3), while the overtones' frequencies are multiples of increasing frequencies from 220 (A3) to 246.94Hz (B3). A listening test indicated that when the overtones are multiples of 246.94Hz, nineteen out of twenty participants perceive A3 and B3 simultaneously -- the $f_0$ and the difference tone. The highest autocorrelation peak (excluding zero lag) is close to A\#3 and does not correspond to any perceived pitch.


\subsection{Source separation}

The analyzes are based either on the original song version or on audio files obtained by source separation using the X-UMX algorithm \citep{stoter2019open}. The algorithms separates the original audio into six categories: drums, guitar, bass, piano, vocals, and `other'.

\subsection{Psychoacoustic weighting}\label{subsec:iso}

Signal analyses in this study incorporate equal-loudness-contour weighting, recognizing that humans are not equally sensitive to all frequencies \citep{fletcher1933loudness}. Various models exist that weight the power spectrum to reflect perception \citep{fletcher1933loudness,robinson1956re}\citep[p.~7]{skovenborg2004evaluation}. One such model is ISO226:2023 \citep{iso2262023}. We apply the ISO226:2023 50-phon equal-loudness contour before spectral analysis to produce results that better reflect what listeners actually hear.

\subsection{Analytical process}\label{subsec:analyticalprocess}

Seventy-two songs originating from seven records released between 2005 and 2023 were imported into a Pro Tools session and subjected to critical listening. The focus of the listening was \textit{pitch}---not the high-level organization of pitches, such as chord sequence identification, but lower-level criteria, such as the complexity and nature of the tones carrying pitch or the compliance of pitches to equal temperament. Observations were sorted into six non-exclusive categories:

\begin{enumerate}[noitemsep]
    \item Presence of high-complexity instrumental tones.
    \item Presence of complex vocal parts (e.g., use of vocoder).
    \item Tuning not consistent with equal temperament.
    \item Presence of continuous pitch trajectories.
    \item Presence of low pitch-strength instruments, e.g., drums.
    \item Presence of flanger/chorus resulting in additional pitch.
\end{enumerate}

The present study focuses on the first category. Thirty-eight of the seventy-two songs were found to involve high-complexity instrumental tones, most found in bass tracks. As a result, while the study's conclusions do not apply to all of Vitalic's productions, they reflect more than half of his discography.

Table~\ref{table:vitalicext} lists the extracts used in this study. The primary criterion for selecting these extracts was practicality. In particular, when using source separation, synthesizer parts are often combined in the `other' category, making it difficult to separate them. Vitalic's bass parts are also frequently split between the `bass' and `other' outputs. A particular case is `No Fun', chosen because it illustrates alternative ways of conveying pitch more clearly than most other examples.

\begin{table*}[t]
\begin{center}

\small

\begin{tabularx}{\textwidth} { 
  | >{\centering\arraybackslash}p{0.21\textwidth} 
  | >{\centering\arraybackslash}p{0.29\textwidth} 
  | >{\centering\arraybackslash}X 
  | >{\centering\arraybackslash}p{0.12\textwidth} 
  | >{\centering\arraybackslash}p{0.08\textwidth} | }

 \hline
    \textbf{Album} & \textbf{Song} & \textbf{Reference} & \textbf{Timing} & \textbf{Section}  \\
\hline
    `OK Cowboy' & `No Fun' & \citep{vitalic2005nofun} &  0'08 to 0'48 & \ref{sec:introduction}, \ref{sec:freedom}\\
\hline   
    `Voyager' & `Use It Or Lose It' & \citep{vitalic2017useitorloseit} &  0'19 to 0'21 & \ref{sec:singletone}, \ref{sub:noisyshifted}\\
\hline
    `Dissid{\ae}nce Episode 1' & `Cosmic Renegade' & \citep{vitalic2022cosmicrenegade} &  2'54 to 3'02 & \ref{sec:singletone}\\
\hline   
    `Rave Age' & `La Mort Sur Le Dance Floor' & \citep{vitalic2012lamortsurledancefloor} &  0'03 to 0'07 & \ref{sub:noisyharm}\\
\hline   
    `Flashmob' & `Station MIR 2099' & \citep{vitalic2009stationmir} &  0'06 to 0'10 & \ref{sub:noisyharm}\\
\hline   
    `Rave Age' & `Next I'm Ready' & \citep{vitalic2012nextimready} &  0'05 to 0'07 & \ref{sub:noisyharm}\\
\hline   
    `OK Cowboy' & `Poney Part 1' & \citep{vitalic2005poneypart1} &  0'53 to 0'55 & \ref{sub:noisyshifted}, \ref{subsec:inharmonicnumber}\\
\hline   
    `Voyager' & `Nozomi' & \citep{vitalic2017nozomi} &  0'42 to 0'44 & \ref{sub:noisystretched}\\
\hline   
    `Dissid{\ae}nce Episode 2' & `Sirens' & \citep{vitalic2022sirens} &  0'09 to 0'13 & \ref{sub:noisystretched}\\
\hline

\end{tabularx}

\caption{Extracts of Vitalic songs used in the study.} 
    \label{table:vitalicext}     
    
\end{center}
\end{table*}


\subsection{Analyses were reviewed by Vitalic}\label{subsec:reviews}

The analysis results were discussed with Vitalic throughout the research and writing process. These discussions focused on three main points:

\begin{enumerate}[label={(\arabic*)}, noitemsep]
    \item Validation. Were the analysis results relevant? Did they highlight important aspects of the musical discourse? 
    \item Production methods. What instrumental processes led to these results?
    \item Artistic intent. Beyond technical aspects, what was the purpose of diverging from equal temperament? Vitalic reviewed and approved all the results in this paper before submission.
\end{enumerate}

\section{Pitch(es) derived from a single quasi-harmonic tone}\label{sec:singletone}


In Vitalic's music, perceived pitches from quasi-harmonic tones may not correspond to the tone's $f_0$, and a single quasi-harmonic tone may convey more than one pitch. Vitalic notes that during production, when the pitch(es) do not correspond to the fundamental, it results in a decorrelation between what is played on the keyboard controller and the resulting melody.

\subsection{Perceived pitch(es) may differ from $f_0$}

\begin{figure}[h!]
  \centering
  \includegraphics[width=\textwidth]{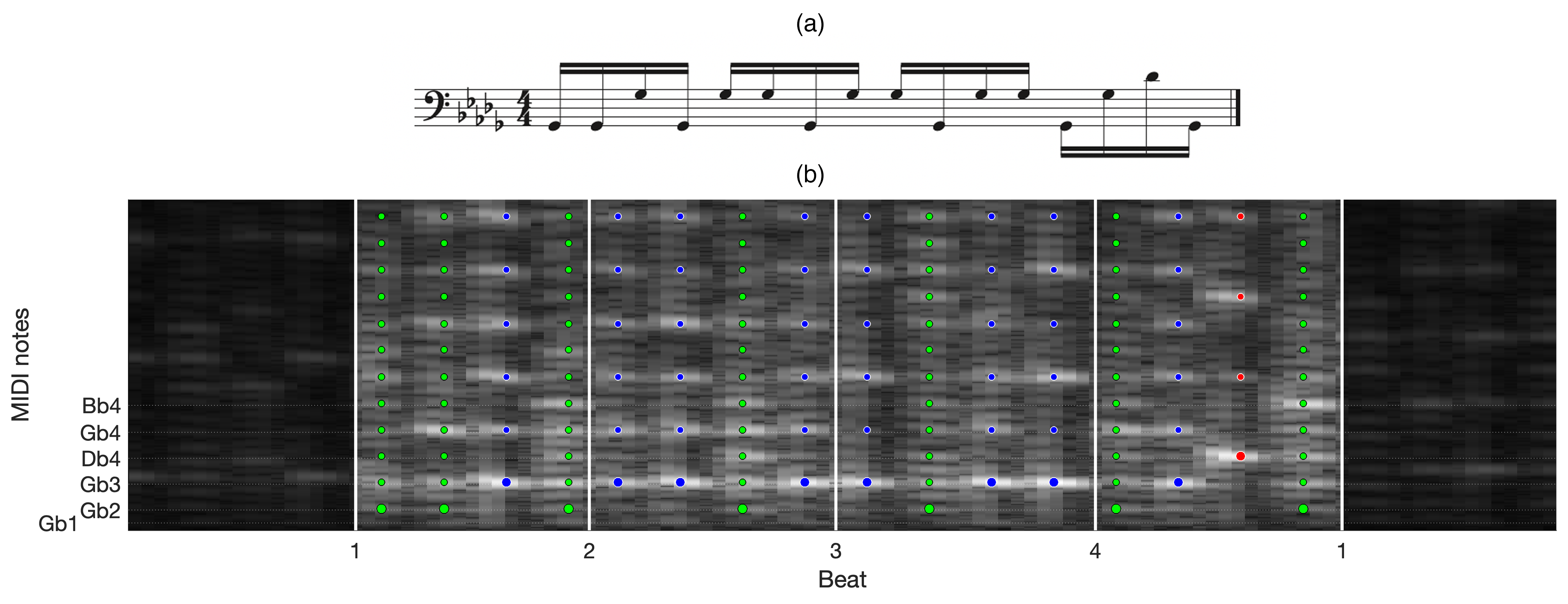}
  \caption{`Cosmic renegade', 2'54 to 3'02, bass (source separation output). See suppl. mat. video 2 for the same figure with synchronised audio. (a) Approximate transcription. (b) STFT, weighted audio. The weighting makes the Gb1 $f_0$ only moderately visible. Scatter plots, formant positions corresponding to the perceived pitches. Yellow, Gb2; blue, Gb3; red, Db4. \\}
\label{fig:cosmicrenegadebass}
\end{figure}

Vitalic deliberately uses quasi-harmonic tones whose perceived pitch(es) do(es) not coincide with the fundamental frequency. As illustrated in Figure~\ref{fig:cosmicrenegadebass}, `Cosmic Renegade' \citep{vitalic2022cosmicrenegade} features a single quasi-harmonic tone with an $f_0$ of Gb1, where the partials' amplitudes are modulated so that the perceived pitches alternate between Gb2, Gb3, and Db4. In this example, the perceived pitch may correspond to the loudest partial (Gb3 and Db4) or the greatest common divisor for a subset of pitches (Gb2).

\subsection{A single tone may prompt several pitches}\label{subsec:severalpitches}

\begin{figure}[H]
  \centering
  \includegraphics[width=1\textwidth]{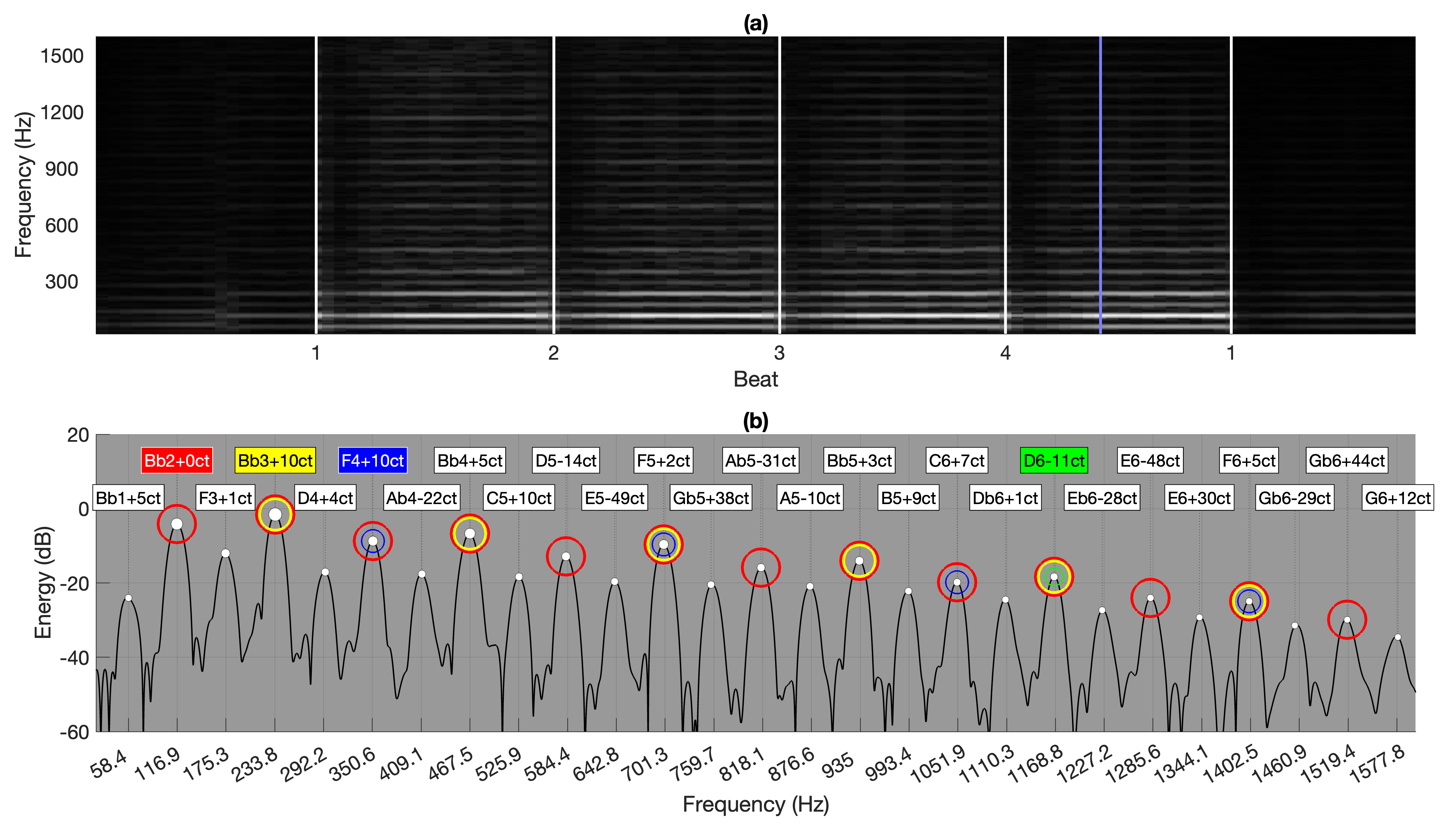}
  \caption{`Use It Or Lose it', 0'19 to 0'21, bass (source separation output). See suppl. mat. video 3 for the same figure with synchronised audio. (a) STFT, \textit{unweighted} audio. (b) FT, \textit{weighted} audio, frame indicated by the blue line in (a). Text boxes, MIDI note corresponding to each partial with deviation in cent. Red and yellow, easily perceivable pitches. Blue and green, two uncertain pitches amongst others. Circles of the same color, same pitches with corresponding overtones.}
\label{fig:useitloseit}
\end{figure}

In Vitalic's music, two or more simultaneous perceivable pitches may stem from a single quasi-harmonic tone. According to Vitalic, the number of perceivable pitches per unique tone is a musical parameter. Figure~\ref{fig:useitloseit} illustrates a bass part from `Use It Or Lose It' \citep{vitalic2017useitorloseit}, where a tone with an $f_0$ of Bb1 results in several perceivable pitches, none of them being Bb1.  

Vitalic notes that the perception of these pitches may be uncertain, with this uncertainty serving as a musical parameter. In `Use It Or Lose It', the easiest pitch to hear is Bb2, followed by Bb3. In the authors' experience, the perception of other pitches depends on the listening system. Using 2021 Mac Book Pro speakers leads to an audible D6. Using AKG MKII K271 headphones, D6 is hardly audible, and F4 becomes more salient. See suppl. mat., video 3, for the methodology that helped identify these pitches.


\newpage

The variable ease with which overtones may result in audible pitches in `Use It Or Lose It' may relate to whether humans can `hear out' individual harmonics. First introduced by \citet{mersenne1636harmonie}, this idea became central to the Ohm--Seebeck dispute, where Ohm argued that only the fundamental partial contributes to a tone's perceived pitch \citep{turner1977ohm}. Experiments indicate humans can hear only the first 5-8 harmonics, though their audibility has been contested \citep{moore2012introduction}. In `Use It Or Lose It', the perceived D6 is the 20\textsuperscript{th} harmonic.

\subsection{Pitch-timbre continuum}

The partials' amplitudes, typically associated with timbre \citep{jensen2002timbre}, influence the perceived pitch(es) in the cases of `Cosmic Renegade' and 'Use It or Lose It'. The influence of the partials' amplitudes on both timbre and pitch reveals a \textit{pitch-timbre continuum}, challenging definitions of timbre that claim it differentiates two sounds with similar other parameters, including pitch \citep{asa2024timbre}.



\section{Inharmonicity}\label{sec:inharmonicity}

Vitalic uses inharmonic tones. Sections~\ref{sub:noisyshifted}--\ref{sub:noisyharm} sorts these tones into three types.

In the most general case, a set of multiple partials consists of partials at random frequencies. A specific case arises when a constraint is applied, requiring that all partials except the lowest-frequency one be equidistant, resulting in what we call type 1. A stricter constraint would require all partials including the lowest-frequency one to be equidistant, resulting in type 2. An even stricter constraint sets the distance between partial frequencies to the frequency of the lowest partial, resulting into harmonic tones, from which type 3 originates.

\subsection{Type 1: (noisy) shifted residue}\label{sub:noisyshifted}

\begin{figure}[h!]
\includegraphics[width=\textwidth]{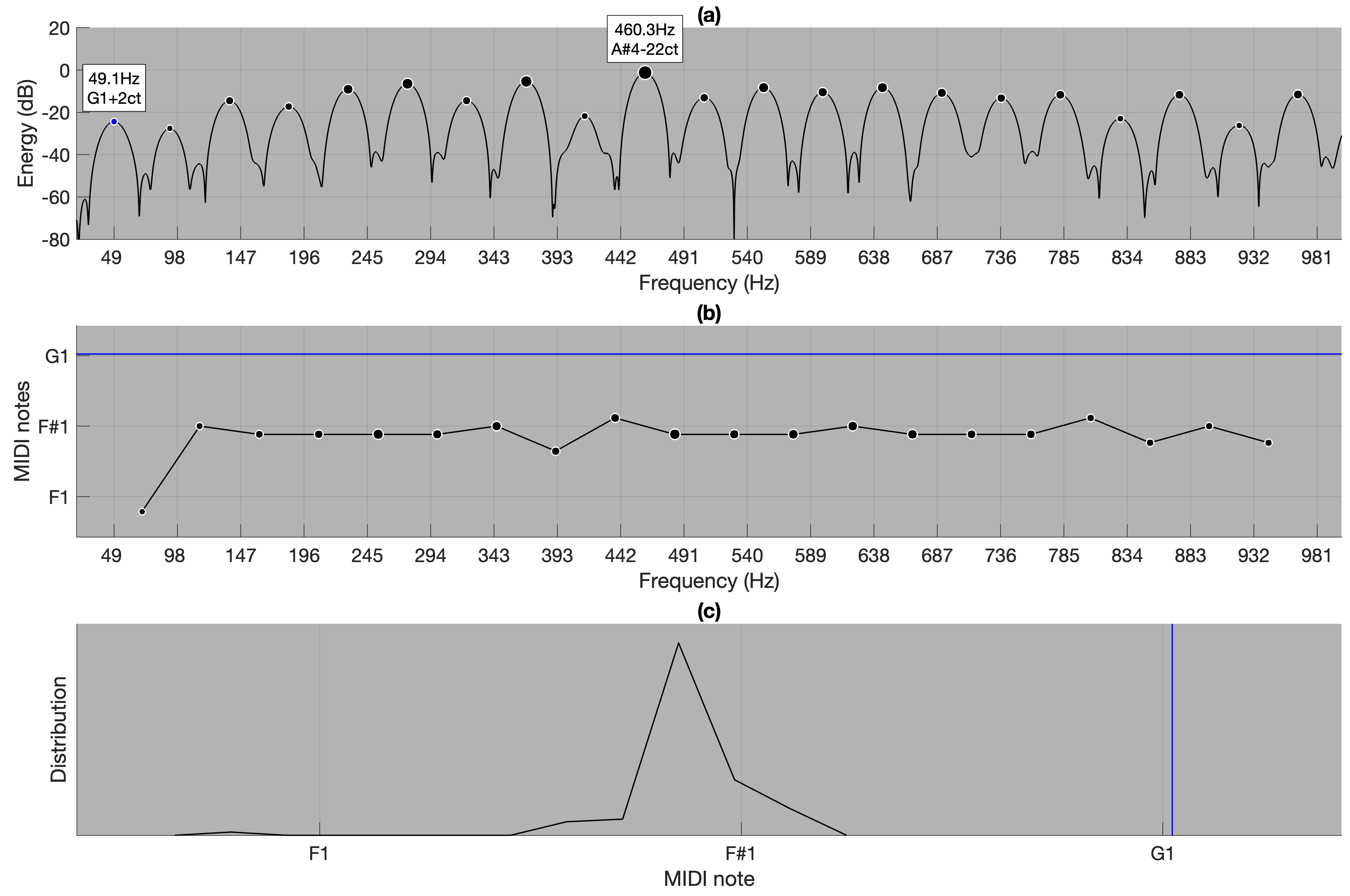}
\caption{`Poney part 1', bass, 0'53 to 0'55, frame 10. See suppl. mat. video 8 for the corresponding audio. (a) Weighted audio, power spectrum. (b) Frequency difference between consecutive partials, expressed as MIDI notes. The blue line shows the lowest partial's frequency. (c) Frequency difference between consecutive partials, distribution. The blue line shows the lowest partial's frequency.}
    \label{fig:Poneyshifted}
\end{figure}

The tones discussed in this section correspond to those described by \citet[p.~365]{schouten1940residue}, \citet[p.~535]{deboer1956pitch}, and \citet[p.~1705]{yost2009pitch}, involving what \citeauthor{yost2009pitch} refers to as a \textit{pitch shift of the residue}. Figure~\ref{fig:Poneyshifted} details a short extract from the bass in `Poney part 1' \citep{vitalic2005poneypart1}.

In such tones, the frequencies of the overtones are `shifted' to the left (lower frequencies) or right (higher frequencies). This `shift' is not the same as transposition: transposition involves shifting partials by an interval (e.g., one semitone), which means multiplying frequencies -- for a semitone, $f \rightarrow f \times (2^{1/12})$. In contrast, frequency `shifting' means adding a fixed value to each frequency -- for a 10Hz shift, $f \rightarrow f + 10$.

Figure~\ref{fig:Poneyshifted} shows an example of overtones shifted towards \textit{lower} frequencies. Due to this shift, the frequency differences between consecutive overtones are centered around a value different from that of the lowest partial. The distribution of differences between partials has one outlier, visible on the left in Figure~\ref{fig:Poneyshifted}(c): the gap between the first two partials. The frequency difference between overtones may vary around a central value, hence the term `noisy'.

An example of overtones shifted towards \textit{higher} frequencies is `Use It or Lose It' \citep{vitalic2017useitorloseit}, suppl. mat. video 9, other representation in Figure~\ref{fig:useitloseit}.

\subsection{Type 2: (noisy) stretched and compressed tones, shifted tones}\label{sub:noisystretched}

\begin{figure}[H]
\includegraphics[width=\textwidth]{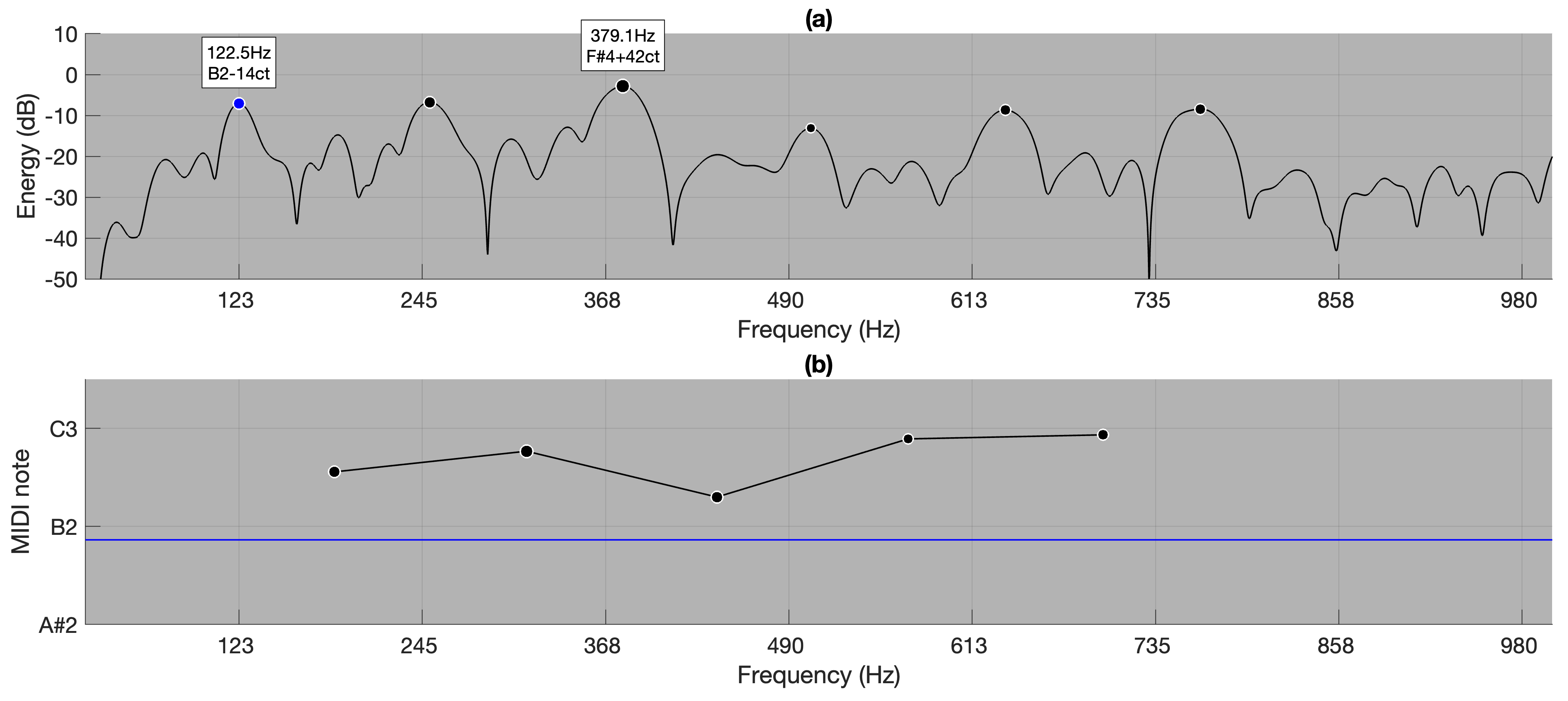}
\caption{`Nozomi', 0'42 to 0'44, frame 15. See suppl. mat. video 10 for the corresponding audio. (a) Weighted audio, power spectrum. (b) Frequency difference between consecutive partials, expressed as MIDI notes. The blue line shows the lowest partial's frequency.} \label{fig:Nozomi}
\end{figure}

Figure~\ref{fig:Nozomi} shows an example of what can be referred to as a `stretched tone'. The corresponding song extract is from `Nozomi' \citep{vitalic2017nozomi}. The frequency differences between consecutive partials exceed the fundamental frequency. Conversely, in compressed tones, the frequency differences between consecutive partials are lower than the fundamental frequency. While in shifted residue tones, the overtones are shifted up/down (e.g., 100, 210, 310, 410... Hz), here, all frequency differences are greater/smaller than the fundamental (e.g., 100, 210, 320, 430... Hz).

Another interpretation of such tones consists in seeing them as `shifted tones', where all partials, including the lowest, are shifted up or down. `No Fun' (Sections~\ref{sec:introduction} and~\ref{sec:freedom}) is based on such tones.

\newpage

`Sirens' \citep{vitalic2022sirens}, suppl. mat. video 10, in addition to involving only odd harmonics, uses `compressed' tones. The frequency differences between consecutive partials, adjusted to the corresponding octave to account for the missing even harmonics, are less than the fundamental frequency. In both cases, the tones remain `noisy', in the sense that the frequency differences between consecutive partials appears to be randomly distributed around a center value.

\subsection{Type 3: noisy harmonicity}\label{sub:noisyharm}

\begin{figure*}
\includegraphics[width=\textwidth]{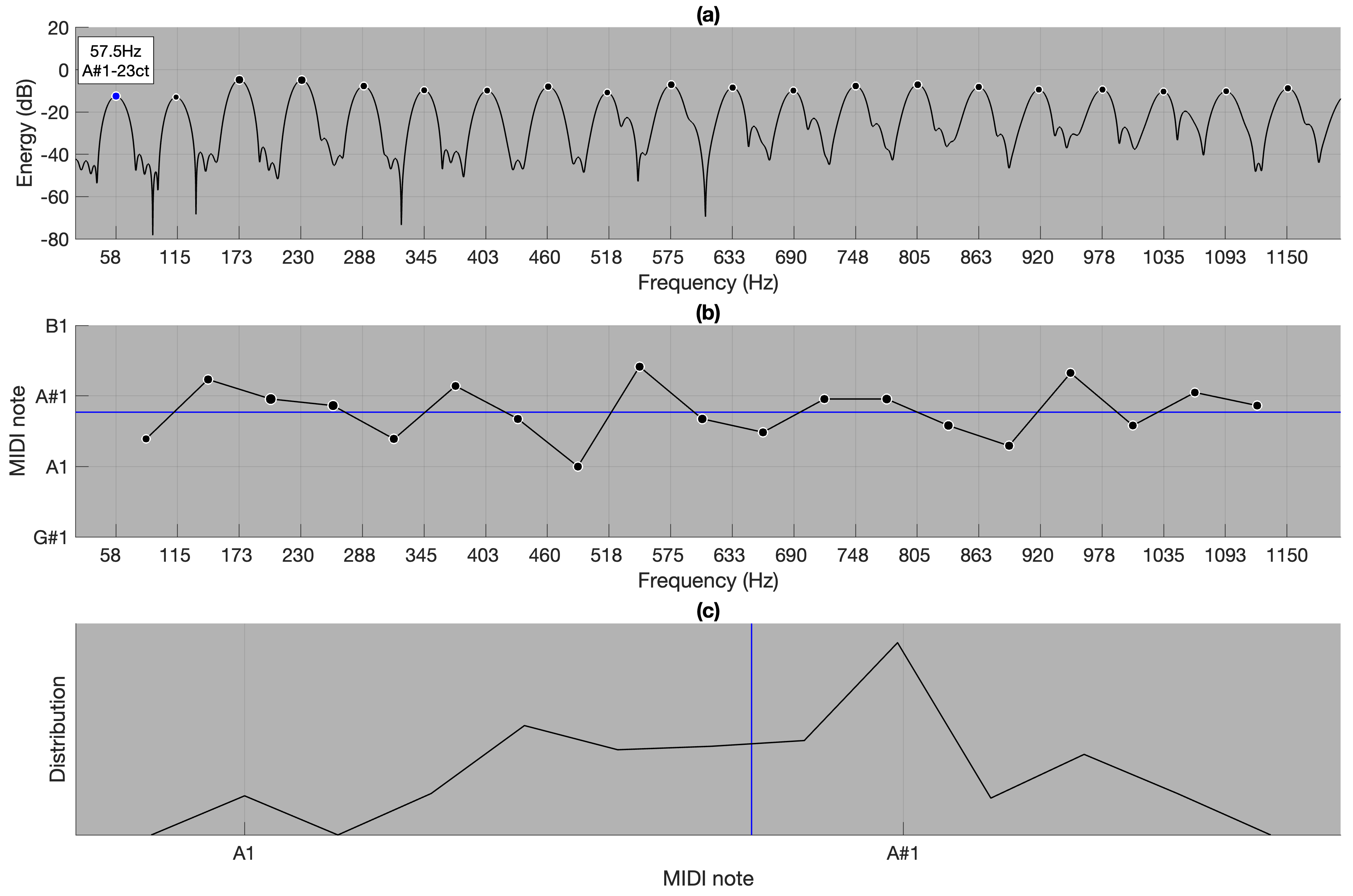}
\caption{`La Mort sur le Dance Floor', 0'03 to 0'07, frame 38. See suppl. mat. video 5 for the corresponding audio. (a) Weighted audio, power spectrum. (b) Frequency difference between consecutive partials, expressed as MIDI notes. The blue line shows the lowest partial's frequency. (c) Frequency difference between consecutive partials, distribution. The blue line shows the lowest partial's frequency.}
    \label{fig:RaveAge}
\end{figure*}

The frequency differences between consecutive partials are distributed around a frequency value close to the fundamental, causing the positions of the spectral peaks to loosely align with the multiples of the fundamental frequency. An example is `La Mort sur le Dance Floor' \citep{vitalic2012lamortsurledancefloor}. As illustrated in Figure~\ref{fig:RaveAge}, the distance to the center frequency appears random, which allows it to be described as `noisy'. The variance of this distribution is a parameter. A track with high variance is `Station MIR 2099' \citep{vitalic2009stationmir}, suppl. mat., video 6.

\begin{figure*}
\includegraphics[width=\textwidth]{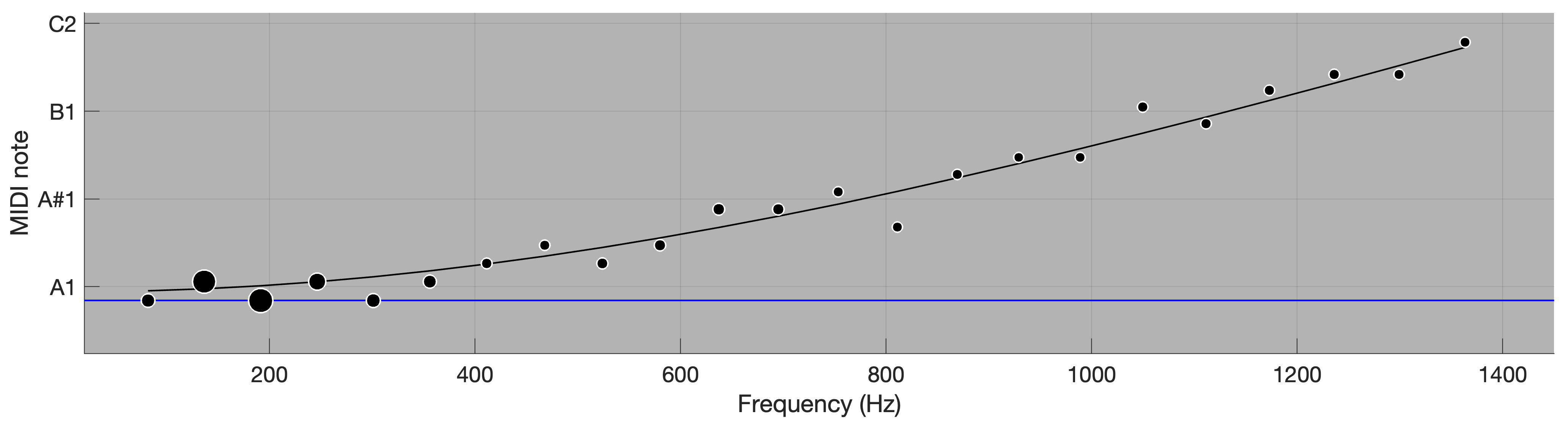}
\caption{Frequency difference between consecutive peaks, piano, A1 sample. The dots show the differences for the first twenty-four partials. The blue horizontal line shows the fundamental frequency. The black line is the best fit for the inharmonicity model. } \label{fig:piano}
\end{figure*}

A subcategory emerges when the frequency differences between consecutive partials are distributed around the octave above the fundamental frequency, due to the absence of even harmonics. Despite this, the spectral peaks still loosely align with the multiples of the fundamental frequency. An example is `Next I'm Ready' \citep{vitalic2012nextimready}, suppl. mat., video 7.

\newpage

\subsection{Inharmonicity in Vitalic basses does not resemble that of piano strings}

Following \citet{schuck1943observations}, \citet{young1952inharmonicity} and \citet{fletcher1964normal}, \citet{heetveld1984string} and \citet{jarvelainen1999audibility} express the frequency of partials in piano tones as $f_n = n f_0 \sqrt{(1+ B n^2)}$, with $B$ being the inharmonicity coefficient. If $B=0$, the tone is harmonic. Figure~\ref{fig:piano} shows the difference between partials for a medium velocity A1 sample from the Native Instruments' `Gentleman' piano library. If 
$f_n = n f_0 \sqrt{(1+ B n^2)}$, then the $n^{th}$ difference between partials can be formulated as $d_n = f_{n+1}-f_n = f_0 ((n+1)  \sqrt{(1+ B (n+1)^2)} - n  \sqrt{(1+ B n^2)})$. For this sample, the model's best fit corresponds to $B=0.00022$, consistent with \citet[p.~173]{heetveld1984string}. None of the tones seen elsewhere in this study comply with this model of inharmonicity.


\subsection{Consequence of inharmonicity on number of perceived pitches}\label{subsec:inharmonicnumber}

Inharmonic complex tones have been described as evoking more than one pitch. If partials are close to harmonic positions, the tone evokes a single pitch, approximately the fundamental of the least-deviating harmonic series \citep[p.~9]{rasch1982perception}. If partials are further from harmonic positions, both the fundamental and other partials dominate \citep{jarvelainen2000effect}. \citet[p.~41]{bregman1996demonstrations} show how one partial becomes audible as inharmonicity increases. 

Therefore, for a given tone, inharmonicity -- i.e., the \textit{position} of partials -- can contribute to the perception of multiple pitches, in addition to the \textit{loudness} of partials (see Section~\ref{subsec:severalpitches}). In Vitalic's discography, we found no clear examples where the perception of multiple pitches from a single tone is solely due to inharmonic relations. In the example from Section~\ref{sub:noisyshifted}, Figure~\ref{fig:Poneyshifted}, for instance, inharmonic relations could have caused the perception of two distinct pitches: F\#1 and G1. However, the partial near G1 is too quiet to be heard. Instead, the perception of multiple pitches likely stems from louder partials, such as the partial near 460Hz (A\#4).



\vspace{1cm}


\section{`No Fun', continued analysis}\label{sec:freedom}

This section elaborates on the main synthesizer part in `No Fun' \citep{vitalic2005nofun}, introduced in Section~\ref{sec:introduction}.

\subsection{The frequency difference between partials is stable}

\begin{figure}[H]
\includegraphics[width=\textwidth]{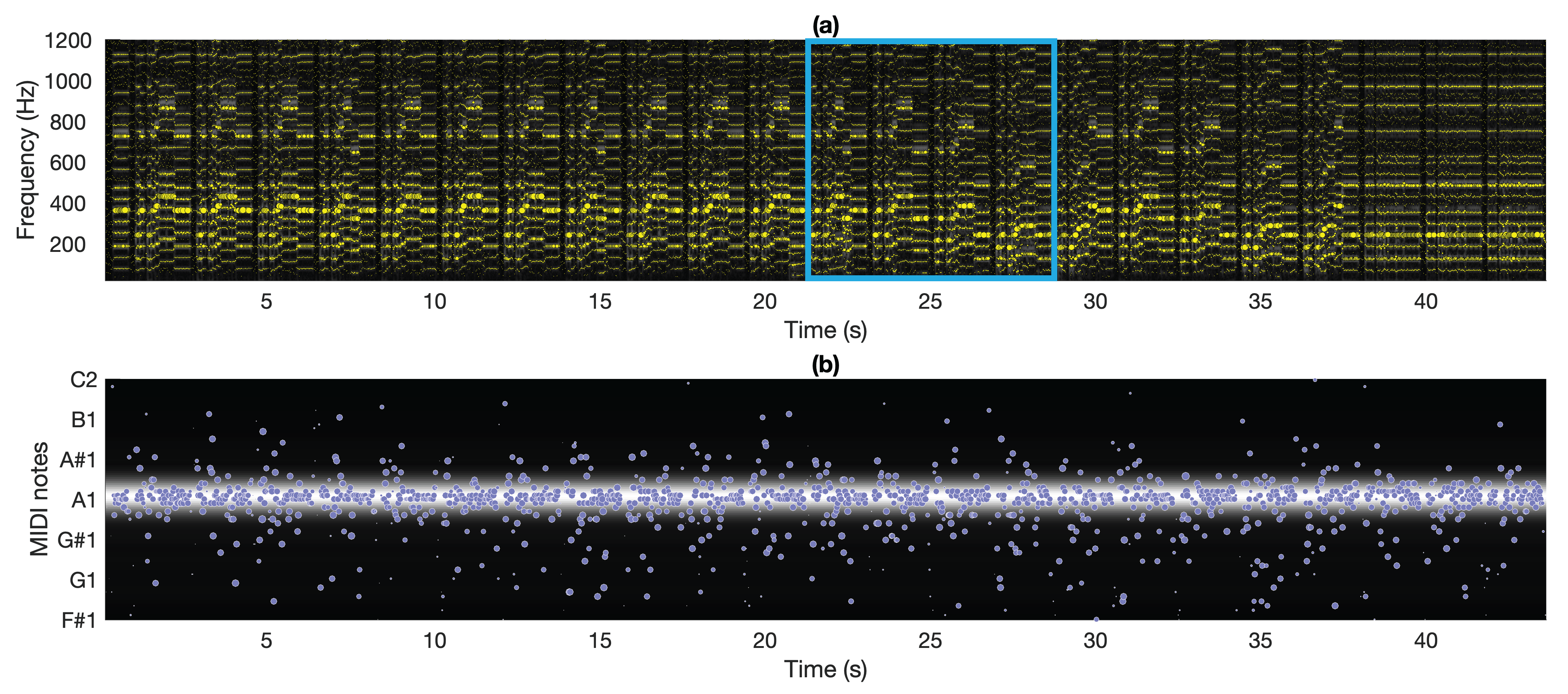}
\caption{`No Fun', 0'08 to 0'48.  Main synthesizer part, from source separation. See suppl. mat. video 12 for the same figure with synchronised audio. (a) STFT of the unweighted audio, with the yellow scatter plot indicating the peaks. A regularity constraint was applied to peak detection. The blue rectangle marks the location of the extract transcribed as a score in Figure~\ref{fig:NoFunWideScore}. (b) Median of the difference between consecutive partials, weighted by their energy. The background shows the distribution of median differences, with a peak at A1 + 5 cents.
} \label{fig:NoFunWideAudio}
\end{figure}

Figure~\ref{fig:NoFunWideAudio} shows the peaks and differences between partials derived from a longer extract of the synthesizer part. Despite the variety of patterns, the frequency difference between partials consistently centers around A1. 

\subsection{A sequence of shifted quasi-harmonic tones}

The main synthesizer in `No Fun' can be seen as a sequence of shifted tones, corresponding to type 2 inharmonicity mentioned in Section~\ref{sub:noisystretched}. The tones are shifted upwards -- which means they can also be interpreted as a series of variably compressed tones.

\begin{figure}[H]
    \includegraphics[width=\textwidth]{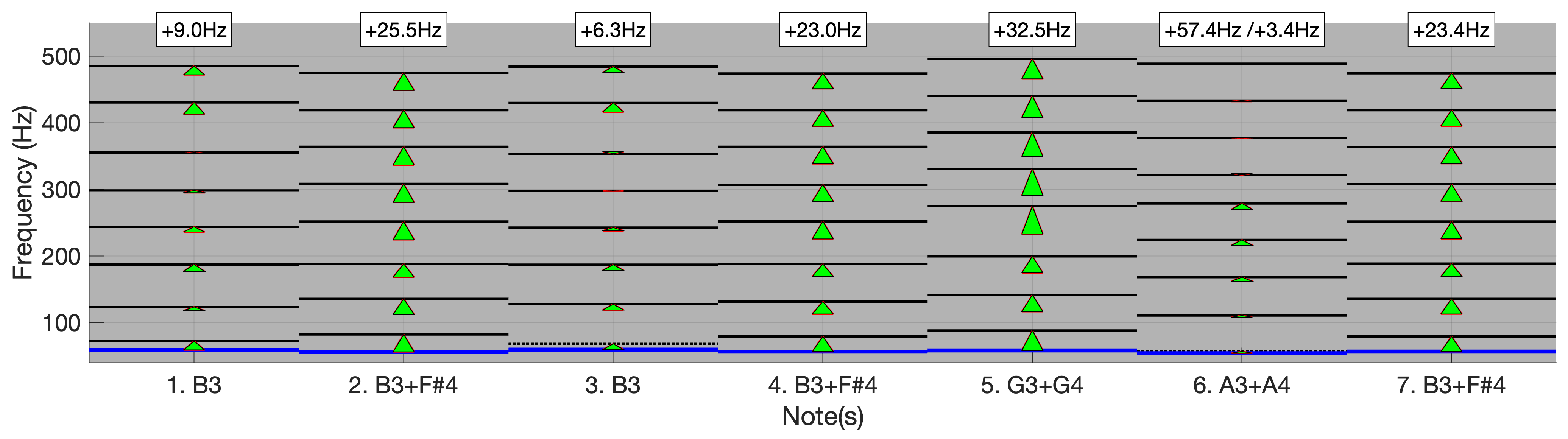}
    \caption{`No Fun', 0'08 to 0'12. Main synthesizer part, from source separation. X-axis: the `notes' correspond to the first pattern in the transcription in Figure~\ref{fig:NoFunWideScore}. Blue lines indicate the mean differences between partials. Solid black lines mark the positions of detected peaks, while dotted black lines indicate the positions of missing peaks near the mean differences between partials. The height of the green triangles represents the frequency difference between multiples of the mean differences between partials and the detected peaks. The framed text displays the mean difference.} \label{fig:NoFunDiffs}
\end{figure}

Figure~\ref{fig:NoFunDiffs} illustrates the upward shift in a sub-sequence of seven `notes' from the synthesizer part. When the shift gets higher than the frequency difference between partials, the frequency difference between partials may be subtracted, and the shift may be interpreted as smaller -- see `note' 6.

\subsection{Perceived pitches: individual partials or partial subsets?}\label{subsec:individualorsubsets}

\begin{figure}[h!]
\includegraphics[width=\textwidth]{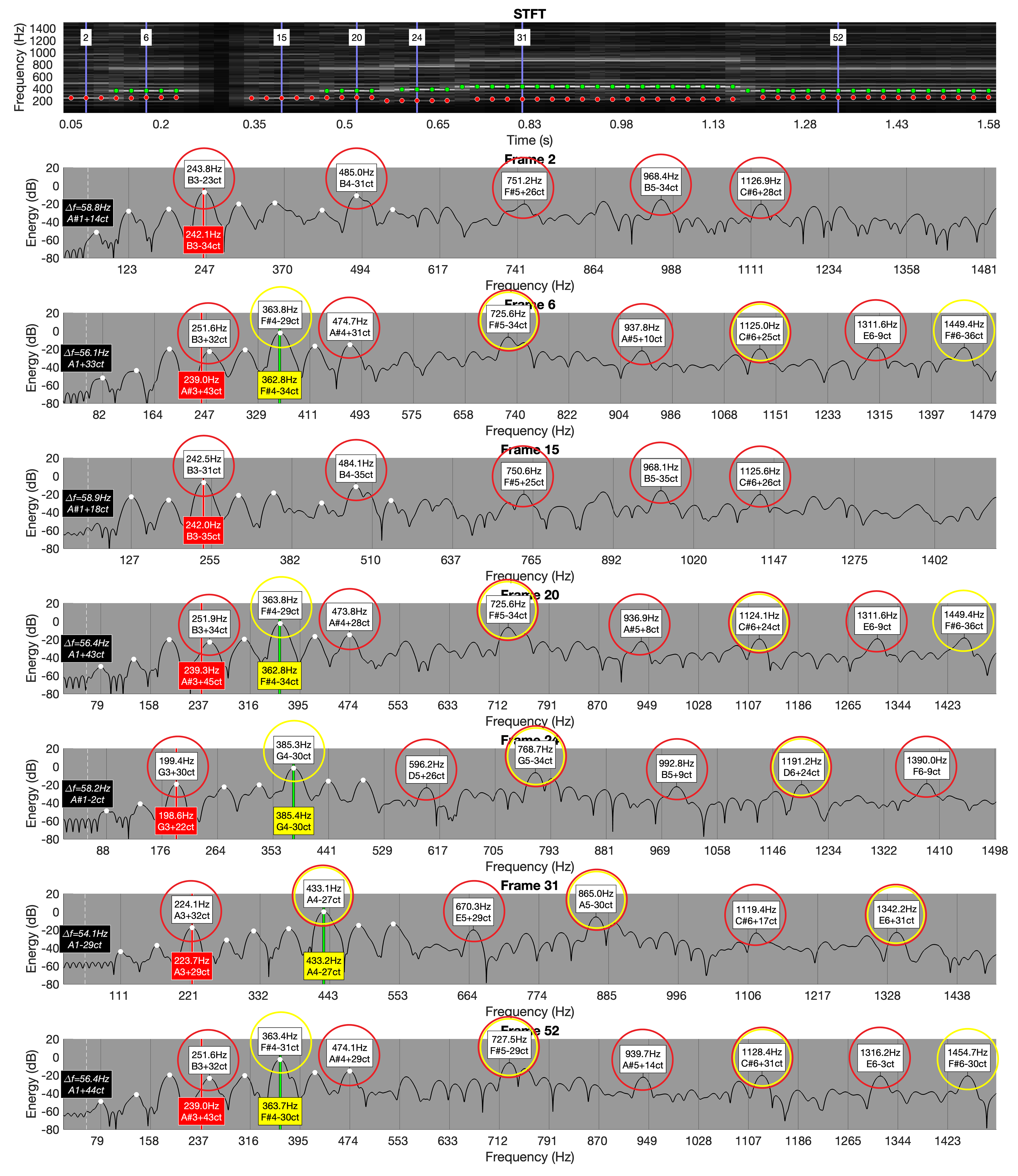}
\caption{`No Fun', 0'08 to 0'12. See suppl. mat. video 13 for the same figure with synchronized audio. Top: weighted audio, STFT. Red and yellow scatter plots, bottom and top melodic lines. Seven lower plots, FT for the frames indicated on the STFT. Red and yellow circles, partials corresponding to the bottom and top melodic lines. Red and yellow text boxes, $f_0$ of the closest harmonic tone. Black text boxes, frequency difference between the partials marked with white dots.} \label{fig:NoFunAll}
\end{figure}

Figure~\ref{fig:NoFunAll} connects Figure~\ref{fig:NoFunWideScore}'s transcription with signal properties, specifically linking the perceived pitches with partials and highlighting the subsets of partials that can be interpreted as forming a quasi-harmonic tone with an $f_0$ corresponding to the perceived pitches.

\clearpage

\clearpage

Although formally linking the tones to perceived pitches is beyond the scope of this paper, we examine whether the perceived pitches in the `No Fun' synthesizer sequence arise from individually audible partials or from the perception of partial subsets. To that effect, we isolate what is referred to as `note 7' in Figure~\ref{fig:NoFunDiffs}, and focus on the F\#4. We process the corresponding audio with four filters:

\begin{enumerate}[label={(\arabic*)}]
    \item Attenuation of the F\#4 $f_0$ (363.4Hz).
    \item Amplification of the F\#4 $f_0$.
    \item Attenuation of the partials identified in Figure~\ref{fig:NoFunAll} as forming a F\#4 quasi-harmonic tone, (363.4Hz, 727.5Hz, 1128.4Hz, and 1454.7Hz).
    \item Amplification of the same partials.
\end{enumerate}

The audio with amplified partials helps confirm the perception of F\#4 (similar process to Section~\ref{subsec:severalpitches}). To determine whether the perception of F\#4 arises from the 363.4Hz partial alone or from the combined partials at 363.4Hz, 727.5Hz, 1128.4Hz, and 1454.7Hz, we compare audio (1) and (3). The perception of F\#4 is stronger in audio (1) than in audio (3), suggesting that it is influenced not only by the $f_0$ but also by the corresponding overtones.  See suppl. mat. video 14 for the audio illustration.

\section{Examples from other sources than Vitalic's music}\label{sec:others}

This section presents examples from sources other than Vitalic's music, demonstrating that similar practices are found elsewhere in popular music.  Sections~\ref{subsec:TR808} and~\ref{subsec:powerchords} illustrate how sounds that are widely used in popular music involve quasi-harmonic tones with individually audible upper partials. Section~\ref{subsec:danger} presents an example of an inharmonic tone similar to those used by Vitalic.

\subsection{TR-808-type bass}\label{subsec:TR808}

The Roland TR-808 Rhythm Composer is a hugely influential drum machine \citep{meyers2003tr,werner2014physically}. A defining sound in hip-hop, the 808's distinctive presets have also become classic in genres like techno, electro, R\&B, and house music \citep{dayal2014tr}. The 808's `long and velvet deep' bass drum \citep{carter1997tr} evolved from a kick drum into a tool for both bass and kick, often carrying the bassline in trap and hip-hop \citep{lavoie2020}. Producers use it to blend drums and bass \citep{Dunn2015,burke2019}.

\begin{figure}[h!]
  \centering
  \includegraphics[width=1\textwidth]{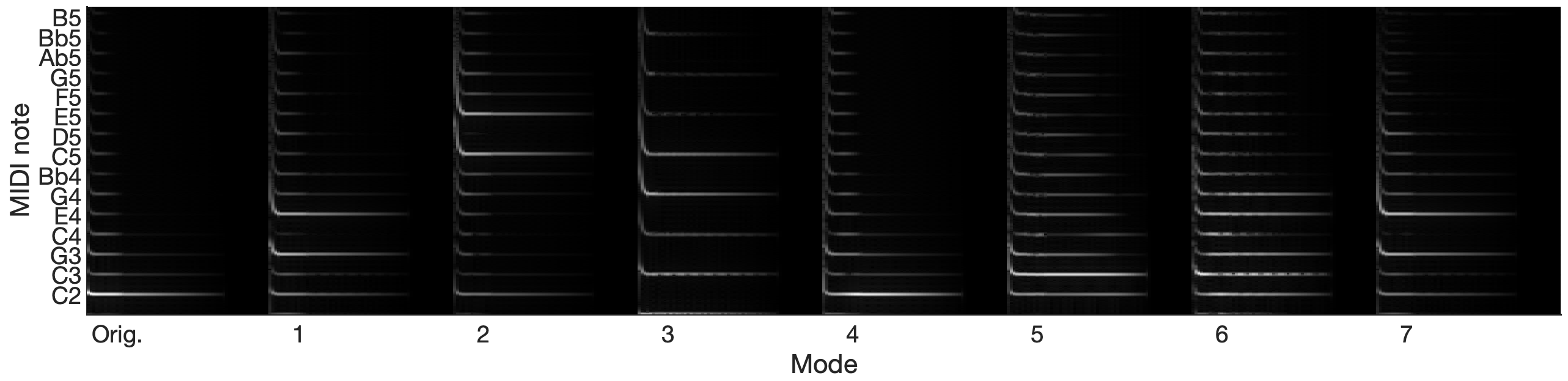}
  \caption{STFT for the Seismic Shock presets (`modes'), weighted audio. Frequencies are expressed as MIDI notes. See suppl. mat. video 15 for the same figure with synchronised audio.} 
\label{fig:seismic}
\end{figure}

Modern computer-based synthesizers may emulate the TR-808 or include TR-808-style libraries. An example is the `808 Woofer Warfare' patch from the Seismic Shock library in Spectrasonics Omnisphere, which produces TR-808-style bass drums. The patch offers seven `modes' (presets). Figure~\ref{fig:seismic} shows that each preset boosts specific upper partials, making certain overtones individually audible. In mode~1, harmonic~5 (two octaves and a major third) is clearly heard. In mode 2, harmonic 10 (three octaves and a major third) is prominent, and in mode~3, harmonic~6 (two octaves and a fifth).


\subsection{Power chords}\label{subsec:powerchords}

\begin{figure}[h]
\includegraphics[width=\textwidth]{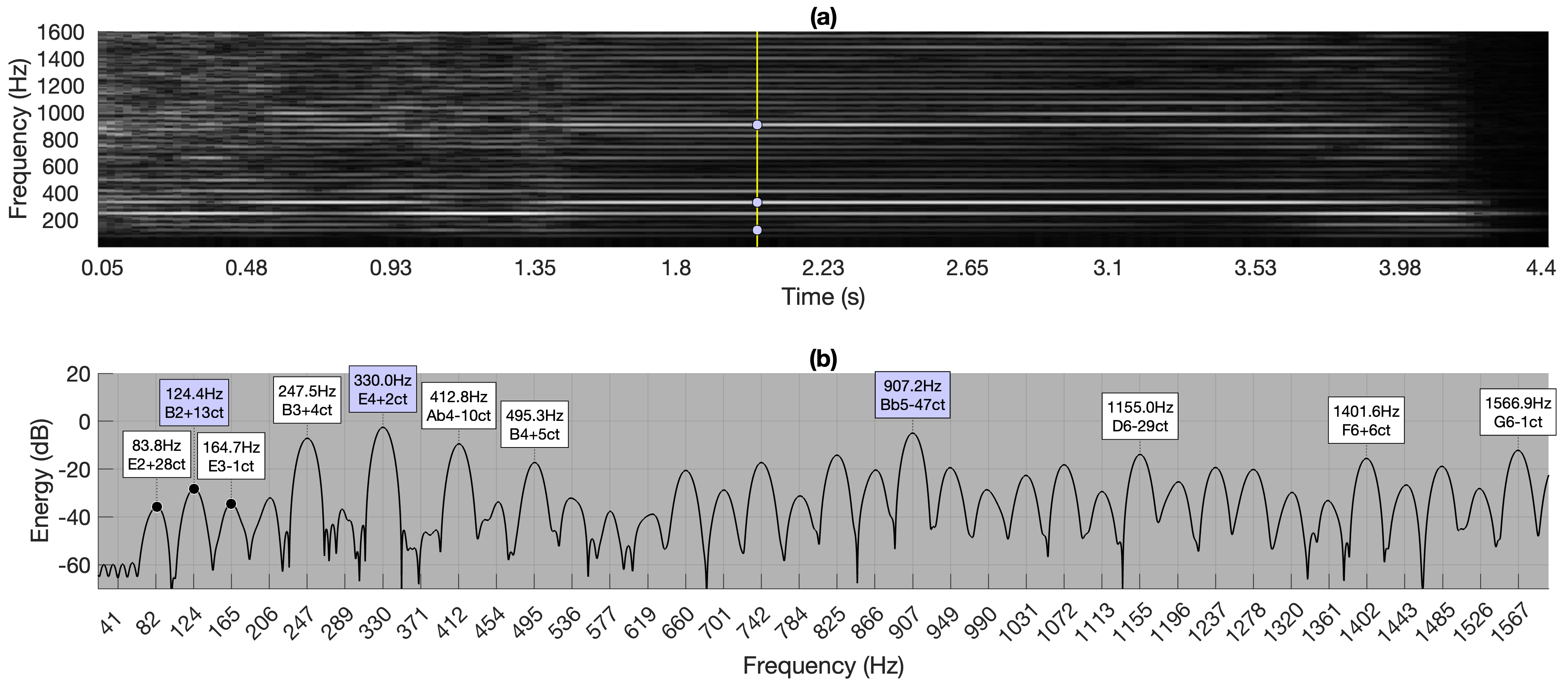}
\caption{Power chord, weighted audio. See suppl. mat. video 16 for the same figure with synchronised audio. (a) STFT. The yellow line indicates the frame analyzed in (b). Blue dots show perceived pitches around this frame, as detailed in (b). (b) FT of the frame shown in yellow. The x-axis is set on the closest harmonic tone. Black dots indicate the played notes, while blue background rectangles represent perceived pitches. See Section~\ref{subsec:severalpitches} for the method used to aid pitch identification.} \label{fig:Pchords}
\end{figure}

 Power chords are a key element of many styles of rock. Figure~\ref{fig:Pchords} shows a harmonic complex tone from a power chord on a guitar, distorted with Native Instrument's Guitar Rig `Rammfire' amp emulation. The played notes are E2, B2 and E3. The result is a quasi-harmonic complex tone whose $f_0$ is 41.2Hz (E1+2cent). The partial corresponding to the $f_0$ is missing. The audible pitches vary across the sample. Near the two-second mark, they are B2, E4, and Bb5. The perception of several pitches from a single quasi-harmonic tone and the decorrelation between played notes and perceived pitches recall Section~\ref{sec:singletone}.


\subsection{Primaal, `Danger', bass track}\label{subsec:danger}

Primaal is a project of the Hyper Music production company \citep{hypermusic}, referred to in \citet{deruty2022development} and \citet{deruty2022melatonin}. In 2022 and 2023, Primaal produced music for brands such as L'Or\'eal, Adidas, Vichy, Honda, GoPro, and Chanel. The producers recognize Trent Reznor, Daft Punk, Kanye West, Skrillex, and Rosalia as influences. 

\begin{figure}[h]
\includegraphics[width=\textwidth]{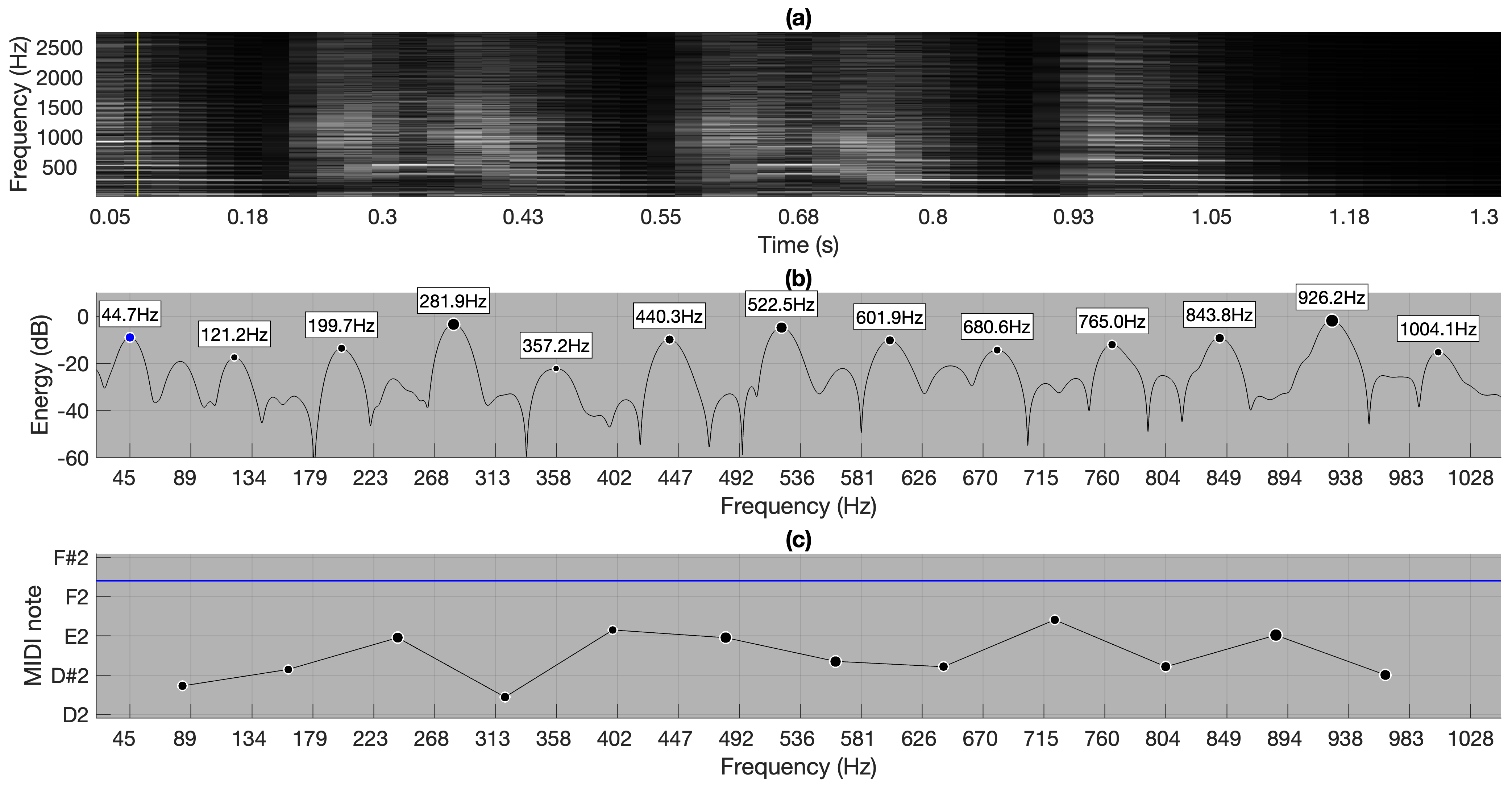}
\caption{Primaal, `Danger', bass track, 0'49 to 0'50, unweighted audio. See suppl. mat. video 17 for the same figure with synchronised audio. (a) STFT. (b) FT, line from (a). The x-axis grid is set on the harmonic positions for the lowest partial. (c) Frequency difference between consecutive partials. The horizontal blue line indicates twice the $f_0$ value from (b) -- twice, to adjust for the preferential use of odd partials.} \label{fig:danger}
\end{figure}

Figure~\ref{fig:danger} shows an extract from one of the bass tracks in Primaal's `Danger', kindly provided by the producers. The bass tone is similar to that of Vitalic's `Nozomi' bass track and belongs to the `noisy stretched tone' category mentioned in Section~\ref{sub:noisystretched}. Several other examples of this type can be found in Primaal's productions.

This track was made the `Subsonic Fracking' patch from Omnisphere, with heavy additional processing: Spectrasonics Seismic Verb, Tape Slammer, Waveshaper, Ring Modulation, with pitch bend and dynamic filtering.


\section{Conclusion}

The French electronic music producer Vitalic makes extensive use of synthesized tones that do not operate on the principle of acoustic resonance. In Section~\ref{sec:introduction}, we introduced the example of the main synthesizer part in Vitalic's `No Fun', composed of a sequence of consecutive inharmonic tones that share the same frequency difference between partials and carry two distinct melodic lines.


\newpage

In Section~\ref{subsec:modeling}, we proposed that linking the signal to perceived pitches in the case of such tones may require a specific approach, involving a simplified form of spectral modeling and a modified form of temporal modeling that includes evaluating the frequency difference between partials instead of relying on autocorrelation.


In Section~\ref{sec:singletone}, we presented several examples from Vitalic's music in which single tones, whether harmonic or inharmonic, lead to the perception of multiple pitches. Vitalic has stated that achieving several distinct, variably clear pitches from a single tone is an intentional process. The perceived pitches correspond to prominent partials, illustrating a pitch-timbre continuum in which the amplitudes of partials affect both timbre and perceived pitch.


In Section~\ref{sec:inharmonicity}, we categorized Vitalic's tones into three distinct types based on inharmonicity: (1) (noisy) shifted residue, (2) (noisy) stretched/compressed tones or shifted tones, and (3) noisy harmonicity. None of these types correspond to the inharmonic tones produced by a piano. We further noted that, in Vitalic's music, the perception of more than one pitch per tone cannot be attributed to inharmonicity but rather to salient partials.


In Section~\ref{sec:freedom}, we elaborated on the example of `No Fun''s main synthesizer, showing that the frequency difference between partials is constant throughout the entire part, and quantifying the shift in the type 2 inharmonic tones found in the part. We suggested that the perceived pitches derive from subsets of the tone's partial and not from individually audible partials. The melodic movements are the combined result of changes in the partial shift and selective amplification of partials.


In Section~\ref{sec:others}, we proposed that Vitalic's approach to the relationship between tones and pitches is rooted in popular music practices. We provided examples of widely used sounds that feature quasi-harmonic tones with individually audible upper partials, as well as an example of an inharmonic tone from other producers that shares common properties with Vitalic's work.



A striking finding, mentioned in Section~\ref{subsec:severalpitches}, resides in the fact that Vitalic uses the number of perceivable pitches per unique tone and the uncertainty surrounding the perception of these pitches as musical parameters. From this perspective, the number of perceived pitches per tone may not be an integer. This is a topic we propose to explore further in other publications.





\newpage
\bibliographystyle{plainnat}
\bibliography{mybib}

@misc{asa2024timbre,
    title={Timbre},
    author={{Acoustical Society of American Standards}},
    year={2024},
    note={\url{https://asastandards.org/terms/timbre/}}
}

@book{bregman1996demonstrations,
  title={Demonstrations to Accompany Bregman's Auditory Scene Analysis},
  author={Bregman, Albert S. and Ahad, Pierre A.},
  year={1996},
  publisher={Cambridge, MA, and London: MIT Press},
  note = {\\\url{https://themusiclab.github.io/bregman-archive/snd/ASA-Demo-Booklet9V4.pdf}}
}

@MISC{burke2019,
  author = {Burke, Gavin},
  title  = {808 Bass and Beyond: Evolution of an Iconic Sound},
  note = {\\\url{https://futureaudioworkshop.com/808-bass-and-beyond-part-i/}},
  year = {2019}
}

@article{carter1997tr,
  title={{Roland TR808 Rhythm Composer (Retro)}},
  author={Carter, Chris},
  journal={Sound on Sound},
  year={1997},
  month ={May},
  note={\\\url{https://www.soundonsound.com/reviews/roland-tr808}}
}

@incollection{dayal2014tr,
  title={{Roland TR-808}},
  author={Dayal, Geeta},
  booktitle={Grove Music Online},
  publisher = {Oxford Music Online},
  year={2014},
  month ={Jan},
  note={\\\url{https://doi.org/10.1093/gmo/9781561592630.article.A2257229}}
}

@article{deboer1956pitch,
  title={Pitch of inharmonic signals},
  author = {De Boer, Egbert},
  sortkey = {Boer, E de},
  journal={Nature},
  volume={178},
  number={4532},
  pages={535--536},
  year={1956},
  publisher={Nature Publishing Group UK London},
  note={\\\url{https://www.nature.com/articles/178535a0}}
}

@article{deruty2022development,
	author = {Deruty, Emmanuel and Grachten, Maarten and Lattner, Stefan and Nistal, Javier and Aouameur, Cyran},
	journal = {Transactions of the International Society for Music Information Retrieval},
	number = {1},
	publisher = {Ubiquity Press},
	title = {On the development and practice of {AI} technology for contemporary popular music production},
	volume = {5},
	year = {2022},
	note = {\\\url{https://transactions.ismir.net/articles/10.5334/tismir.100}}
}

@inproceedings{deruty2022melatonin,
  author       = {Deruty, Emmanuel and
                  Grachten, Maarten},
  title        = {``{M}elatonin'': A Case Study on {AI}-induced Musical 
                   Style},
  booktitle    = {{Proceedings of the 3rd Conference on AI Music 
                   Creativity}},
  year         = 2022,
  publisher    = {AIMC},
  month        = sep,
  note = {\\\url{https://doi.org/10.5281/zenodo.7088302}}
  }

@article{drugman2018traditional,
  title={Traditional machine learning for pitch detection},
  author={Drugman, Thomas and Huybrechts, Goeric and Klimkov, Viacheslav and Moinet, Alexis},
  journal={IEEE Signal Processing Letters},
  volume={25},
  number={11},
  pages={1745--1749},
  year={2018},
  publisher={IEEE},
  note={\\\url{https://doi.org/10.1109/LSP.2018.2874155}}
}

@MISC{Dunn2015,
  author = {Dunn, Alexander},
  title  = {808 (documentary film)},
  note= {\url{https://youtu.be/KClqn0oN1lY}},
  year = {2015}
}

@article{fletcher1933loudness,
	author = {Fletcher, Harvey and Munson, Wilden A},
	journal = {Bell System Technical Journal},
	number = {4},
	pages = {377--430},
	publisher = {Wiley Online Library},
	title = {Loudness, its definition, measurement and calculation},
	volume = {12},
	year = {1933},
	note = {\\\url{https://doi.org/10.1002/j.1538-7305.1933.tb00403.x}}
}

@article{fletcher1964normal,
  title={Normal vibration frequencies of a stiff piano string},
  author={Fletcher, Harvey},
  journal={The Journal of the Acoustical Society of America},
  volume={36},
  number={1},
  pages={203--209},
  year={1964},
  publisher={Acoustical Society of America},
  note={\url{https://doi.org/10.1121/1.1918933}}
}

@article{goldstein1973optimum,
  title={An optimum processor theory for the central formation of the pitch of complex tones},
  author={Goldstein, Julius L},
  journal={The Journal of the Acoustical Society of America},
  volume={54},
  number={6},
  pages={1496--1516},
  year={1973},
  publisher={Acoustical Society of America},
  note={\\\url{https://doi.org/10.1121/1.1914448}}
}

@article{heetveld1984string,
  title={String inharmonicity and piano tuning},
  author={Rasch, Rudolf A and Heetvelt, Vincent},
  journal={Music Perception},
  volume={3},
  number={2},
  pages={171--189},
  year={1985},
  publisher={University of California Press},
  note={\url{https://doi.org/10.2307/40285331}}
}

@book{helmoltz1885sensations,
  title={On the sensations of tone as a physiological basis for the theory of music},
  author={Helmholtz, Hermann Ludwig Ferdinand von},
  publisher={Longmans, Green, and Co.},
  year={1885},
  note={Translated by Alexander John Ellis. \\\url{https://archive.org/details/onsensationsofto00helmrich}}
}

@misc{hypermusic,
  author = {{Hyper Music}},
  title = {Hyper Music},
  year = {2024},
  note = {\url{https://www.hyper-music.com/}}
}

@techreport{iso2262023,
    author = {ISO},
    type = {Standard},
    key = {ISO 226:2023},
    year = {2023},
    title = {{Normal equal-loudness level contours-ISO 226: 2023}},
    address = {Geneva, Switzerland},
    institution = {International Organization for Standardization},
    note = {\url{https://www.iso.org/standard/83117.html}}
}

@inproceedings{jarvelainen1999audibility,
  title={Audibility of Inharmonicity in String Instrument Sounds, and Implications to Digital Sound Synthesis.},
  author={J{\"a}rvel{\"a}inen, Hanna and V{\"a}lim{\"a}ki, Vesa and Karjalainen, Matti},
  booktitle={	Proceedings of the 25th International Computer Music Conference, ICMC 1999, Beijing, China},
  year={1999},
  note={\\\url{http://hdl.handle.net/2027/spo.bbp2372.1999.419}}
}

@inproceedings{jarvelainen2000effect,
  title={The effect of inharmonicity on pitch in string instrument sounds.},
  author={J{\"a}rvel{\"a}inen, Hanna and Verma, Tony S and V{\"a}lim{\"a}ki, Vesa},
  booktitle={Proceedings of the 26th International Computer Music Conference, ICMC 2000, Berlin, Germany},
  year={2000},
  note={\url{http://hdl.handle.net/2027/spo.bbp2372.2000.176}}
}

@article{jensen2002timbre,
  title={The timbre model},
  author={Jensen, Kristoffer},
  journal={Journal of the Acoustical Society of America},
  volume={112},
  number={5},
  pages={2238--2238},
  year={2002},
  publisher={[New York: Acoustical Society of America]},
  note = {\url{https://www.musanim.com/pdf/JensenTimbreModel.pdf}}
}

@inproceedings{kim2018crepe,
  title={Crepe: A convolutional representation for pitch estimation},
  author={Kim, Jong Wook and Salamon, Justin and Li, Peter and Bello, Juan Pablo},
  booktitle={2018 IEEE International Conference on Acoustics, Speech and Signal Processing (ICASSP)},
  pages={161--165},
  year={2018},
  organization={IEEE},
  note={\url{https://doi.org/10.1109/ICASSP.2018.8461329}}
  }

@MISC{lavoie2020,
  author = {Lavoie, Alex},
  title  = {What is an 808? 7 Ways to Make Huge 808 Kicks},
  note = {\\\url{https://blog.landr.com/what-is-an-808/}},
  year = {2020},
  month = {Sep.}
}

@article{licklider1951duplex,
  title={A duplex theory of pitch perception},
  author={Licklider, Joseph Carl Robnett},
  journal={The Journal of the Acoustical Society of America},
  volume={23},
  number={1, Supplement},
  pages={147--147},
  year={1951},
  publisher={AIP Publishing},
  note = {\url{https://doi.org/10.1121/1.1917296}}
}

@inproceedings{mauch2010approximate,
  author       = {Matthias Mauch and
                  Simon Dixon},
  title        = {Approximate Note Transcription for the Improved 
                   Identification of Difficult Chords},
  booktitle    = {{Proceedings of the 11th International Society for 
                   Music Information Retrieval Conference}},
  year         = 2010,
  pages        = {135-140},
  publisher    = {ISMIR},
  month        = sep,
  venue        = {Utrecht, Netherlands},
  note          = {\url{https://doi.org/10.5281/zenodo.1416598}}
}

@book{mersenne1636harmonie,
    title={Harmonie Universelle},
    author={Mersenne, Marin},
    year={1636},
    publisher={Paris: chez S\'ebastien Cramoisy},
   note={\\\url{https://imslp.org/wiki/Harmonie_universelle_(Mersenne,_Marin)}}
}

@misc{meyers2003tr,
  title={{Roland TR-808 Rhythm Composer}},
  author={Meyers, Owen},
  publisher={McGill University},
  year={2003},
  note={\\\url{https://www.academia.edu/download/105429503/tr_808.pdf}}
}

@article{moore1986thresholds,
  title={Thresholds for hearing mistuned partials as separate tones in harmonic complexes},
  author={Moore, Brian CJ and Glasberg, Brian R and Peters, Robert W},
  journal={The Journal of the Acoustical Society of America},
  volume={80},
  number={2},
  pages={479--483},
  year={1986},
  publisher={Acoustical Society of America},
  note={\url{https://doi.org/10.1121/1.394043}}
}

@book{moore2012introduction,
  title={An introduction to the psychology of hearing},
  author={Moore, Brian CJ},
  year={2012},
  publisher={Emerald Group Publishing Limited},
  note={ISBN 9780125056274}
}

@misc{phares2024vitalic,
    title={Vitalic Biography},
    author={Phares, Heather},
    year={2024},
    howpublished={Allmusic.com},
    note={\\\url{https://www.allmusic.com/artist/vitalic-mn0000220024#biography}}
}

@article{rabiner1976comparative,
  title={A comparative performance study of several pitch detection algorithms},
  author={Rabiner, Lawrence and Cheng, Md and Rosenberg, A and McGonegal, C},
  journal={IEEE Transactions on Acoustics, Speech, and Signal Processing},
  volume={24},
  number={5},
  pages={399--418},
  year={1976},
  publisher={IEEE},
  note={\url{https://doi.org/10.1109/TASSP.1976.1162846}}
}

@incollection{rasch1982perception,
	author = {Rasch, Rudolf and Plomp, Reinier},
	booktitle = {The Psychology of Music},
	editor = {Diana Deutsch},
	pages = {89--112},
	publisher = {Academic Press},
	title = {The perception of musical tones},
	year = {1982},
        note={\\\url{https://doi.org/10.1016/B978-012213564-4/50005-6}}
}

@article{robinson1956re,
  title={A re-determination of the equal-loudness relations for pure tones},
  author={Robinson, Derek W and Dadson, R So},
  journal={British Journal of Applied Physics},
  volume={7},
  number={5},
  pages={166},
  year={1956},
  publisher={IOP Publishing},
  note={\url{https://doi.org/10.1088/0508-3443/7/5/302}}
}

@inproceedings{schouten1940residue,
  title={The residue, a new component in subjective sound analysis},
  author={Schouten, John F},
  booktitle={Proceedings of the Koninklijke Nederlandse Akademie van Wetenschappen, 1940},
  volume={43},
  pages={356--365},
  year={1940},
  publisher={	North Holland Publ. Co.: Amsterdam},
  note={\url{https://dwc.knaw.nl/DL/publications/PU00017418.pdf}}
}

@article{schuck1943observations,
  title={Observations on the vibrations of piano strings},
  author={Schuck, OH and Young, RW},
  journal={The Journal of the Acoustical Society of America},
  volume={15},
  number={1},
  pages={1--11},
  year={1943},
  publisher={Acoustical Society of America},
  note={\url{https://doi.org/10.1121/1.1916221}}
}

@inproceedings{skovenborg2004evaluation,
  title={Evaluation of different loudness models with music and speech material},
  author={Skovenborg, Esben and Nielsen, Soren H},
  booktitle={Audio Engineering Society Convention 117},
  year={2004},
  organization={Audio Engineering Society},
  note={\\\url{https://aes2.org/publications/elibrary-page/?id=12891}}
}

@article{stoter2019open,
  title={Open-unmix-a reference implementation for music source separation},
  author={St{\"o}ter, Fabian-Robert and Uhlich, Stefan and Liutkus, Antoine and Mitsufuji, Yuki},
  journal={Journal of Open Source Software},
  volume={4},
  number={41},
  pages={1667},
  year={2019},
  note={\url{https://doi.org/10.21105/joss.01667}}
}

@article{turner1977ohm,
  title={The {Ohm-Seebeck dispute}, {Hermann von Helmholtz}, and the origins of physiological acoustics},
  author={Turner, R Steven},
  journal={The British Journal for the History of Science},
  volume={10},
  number={1},
  pages={1--24},
  year={1977},
  publisher={Cambridge University Press},
  note={\\\url{https://www.jstor.org/stable/pdf/4025578.pdf}}
}

@misc{vitalic2005nofun,
  author = {Vitalic},
  title = {No Fun [song]},
  howpublished = {Album: OK Cowboy},
  year = {2005},
  note = {Record Label: Different Records / PIAS. \\\url{https://youtu.be/f8K7Pbxhgps}}
}

@misc{vitalic2005poneypart1,
  author = {Vitalic},
  title = {Poney Part 1 [song]},
  howpublished = {Album: OK Cowboy},
  year = {2005},
  note = {Record Label: Different Records / PIAS. \url{https://youtu.be/9jj4kf7SleA}}
}

@misc{vitalic2009stationmir,
  author = {Vitalic},
  title = {Station {MIR} 2099 [song]},
  howpublished = {Album: Flashmob},
  year = {2009},
  note = {Record Label: Different Records / PIAS. \url{https://youtu.be/RVX-J1yt_vA}}
}

@misc{vitalic2012lamortsurledancefloor,
  author = {Vitalic},
  title = {La Mort sur le Dance Floor [song]},
  howpublished = {Album: Rave Age},
  year = {2012},
  note = {Record Label: Different Records / PIAS. \url{https://youtu.be/jhsIyAVz1eI}}
}

@misc{vitalic2012nextimready,
  author = {Vitalic},
  title = {Next {I}'m Ready [song]},
  howpublished = {Album: Rave Age},
  year = {2012},
  note = {Record Label: Different Records / PIAS. \url{https://youtu.be/FhOfNgDNia8}}
}

@misc{vitalic2017eternity,
  author = {Vitalic},
  title = {Eternity [song]},
  howpublished = {Album: Voyager},
  year = {2017},
  note = {Record Label: Clivage Music / Universal. \\\url{https://youtu.be/u9U_xOIdz_s}}
}

@misc{vitalic2017nozomi,
  author = {Vitalic},
  title = {Nozomi [song]},
  howpublished = {Album: Voyager},
  year = {2017},
  note = {Record Label: Clivage Music / Universal. \\\url{https://youtu.be/CzHfyOGWDSA}}
}

@misc{vitalic2017useitorloseit,
  author = {Vitalic},
  title = {Use it or Lose it [song]},
  howpublished = {Album: Voyager},
  year = {2017},
  note = {Record Label: Clivage Music / Universal. \url{https://youtu.be/J4IEHrFI07w}}
}

@misc{vitalic2022cosmicrenegade,
  author = {Vitalic},
  title = {Cosmic Renegade [song]},
  howpublished = {Album: Dissid\ae nce (Episode 1)},
  year = {2022},
  note = {Record Label: Clivage Music. \url{https://youtu.be/Mis0A-ZAVwM}}
}

@misc{vitalic2022sirens,
  author = {Vitalic},
  title = {Sirens [song]},
  howpublished = {Album: Dissid\ae nce (Episode 2)},
  year = {2022},
  note = {Record Label: Clivage Music. \\\url{https://youtu.be/_z49a3bhXT8}}
}

@misc{vitalic2023anditgoeslike,
  author = {{Vitalic}},
  title = {And it Goes Like [song]},
  howpublished = {Album: Confess EP},
  year = {2023},
  note = {Record Label: Citizen Records / Virgin Music France. \url{https://youtu.be/n97mOW3d1n8}}
}

@inproceedings{werner2014physically,
  author       = {Werner, Kurt James and Abel, Jonathan S and Smith III, Julius O},
  title        = {A Physically-Informed, Circuit-Bendable, Digital Model of the {Roland TR-808} Bass Drum Circuit},
  booktitle    = {{Proceedings of the 17th International Conference on Digital Audio Effects (DAFx-14)}},
  year         = 2014,
  publisher    = {International Audio Laboratories Erlangen},
  month        = sep,
  pages={159--166},
  venue        = {Erlangen, Germany},
  note={\url{https://www.dafx14.fau.de/papers/dafx14_kurt_james_werner_a_physically_informed,_ci.pdf}}
}

@incollection{Wertheimer1938,
    author = {M. Wertheimer},
    title = {Laws of organization in perceptual forms},
    booktitle = {A Source Book of Gestalt Psychology},
    publisher = {Kegan Paul, Trench, Trubner \& Company},
    year = 1938,
    pages = "71--88",
    editor = "W. D. Ellis",
    note = {\\\url{https://doi.org/10.1037/11496-005}}
}

@article{yost2009pitch,
  	title={Pitch perception},
  	author={Yost, William A},
  	journal={Attention, Perception, \& Psychophysics},
  	volume={71},
  	number={8},
  	pages={1701--1715},
  	year={2009},
  	publisher={Springer},
  	note={\url{https://doi.org/10.3758/APP.71.8.1701}}
}

@article{young1952inharmonicity,
  title={Inharmonicity of plain wire piano strings},
  author={Young, Robert W},
  journal={The Journal of the Acoustical Society of America},
  volume={24},
  number={3},
  pages={267--273},
  year={1952},
  publisher={Acoustical Society of America},
  note={\url{https://doi.org/10.1121/1.1906888}}
}

\end{document}